\documentclass[usenatbib,preprint,onecolumn]{mn2e}
\usepackage{graphicx}
\usepackage {natbib,aas_macros}
\usepackage {bm}
\usepackage {amsmath,amssymb}
\usepackage {setspace}
\newcommand{\cm}{\,{\rm g~ cm}^{-3} } 
\newcommand{\cmm}{\,{\rm  cm}^{-3} }

\title{An explicit scheme for ohmic dissipation with smoothed particle magneto-hydrodynamics}

\author[Tsukamoto et al]{Yusuke Tsukamoto$^{1}$, Kazunari
Iwasaki$^{1}$ and  Shu-ichiro Inutsuka$^{1}$ \\
$^1$Department of Physics, Nagoya University, Furo-cho, Chikusa-ku, Nagoya, Aichi, Japan  
}

\begin{document}
\maketitle

\begin{abstract}
In this paper, we present an explicit scheme for Ohmic dissipation
with smoothed particle magneto-hydrodynamics (SPMHD). 
We propose a SPH discretization of Ohmic dissipation and
solve Ohmic dissipation part of induction equation with
the super-time-stepping method (STS) which allows us to take a longer
time-step than Courant-Friedrich-Levy stability condition. 
Our scheme is second-order accurate in space
and first-order accurate in time. 
Our numerical experiments show that optimal choice of 
the parameters of STS for Ohmic dissipation of
SPMHD is $\nu_{{\rm sts}}\sim0.01$ and $N_{{\rm sts}}\sim 5$.
\end{abstract}


\section{Introduction}
Magnetic field plays an important role in various astrophysical problems.
In star formation processes, magnetic field
changes the formation and evolution of protostars, disks and jets.
\citep[e.g.,][]{2004MNRAS.348L...1M,2004ApJ...616..266M,2007PASJ...59..905M,2010ApJ...718L..58I,
2011ApJ...729...42M}. 
Until recently, these interesting phenomena in collapsing magnetized cloud core have been 
investigated with nested-grid code or adaptive-mesh-refinement code.

Smoothed particle hydrodynamics (SPH) is a suitable numerical scheme for
the protostellar collapse simulations because of its adaptive nature at
high density region and several authors have investigated the formation
and evolution of protostar and disk in molecular cloud core.
\citep[e.g.,][]{1998ApJ...508L..95B,2011MNRAS.416..591T,2013MNRAS.428.1321T,2011ApJ...730...32S}.
In spite of the importance of magnetic field, however, most of the simulations with SPH do not include magnetic
field because, until recently, robust magneto-hydrodynamics (MHD)
schemes for SPH have not been developed.

Recently, several authors proposed robust smoothed particle 
magneto-hydrodynamics (SPMHD) schemes.
\citet{2012JCoPh.231.7214T} proposed a SPMHD scheme with the hyperbolic
divergence cleaning method \citep{2002JCoPh.175..645D}, which is
originally proposed by \citet{2005MNRAS.364..384P}. They improved the original
method of \citet{2005MNRAS.364..384P} by changing discretization 
forms for $\nabla \cdot \bm{B}$ and $\nabla \phi$. 
With this method, they successfully simulated protostellar collapse and formation of jets
\citep[see, also][]{2012MNRAS.423L..45P}.

\citet{2011MNRAS.418.1668I} proposed a SPMHD scheme based on Godunov SPH
(GSPH) proposed by \citet{2002JCoPh.179..238I}. We refer to their method
as Godunov smoothed particle magneto-hydrodynamics (GSPMHD).
Instead of the artificial dissipation terms which are used in
\citet{2005MNRAS.364..384P}, they use a solution of a non-linear Riemann
problem with magnetic pressure and the Method of Characteristics to
calculate the interactions between SPH particles. This method significantly reduces the
numerical diffusion (compare, figure 2 of \citealt{2011MNRAS.418.1668I} and
figure 6 of \citealt{2005MNRAS.364..384P}).
They also have developed hyperbolic divergence cleaning method for
GSPMHD ((Iwasaki \& Inutsuka in prep) and successfully simulated formation
of jets (Iwasaki in prep).

In the previous studies about star formation processes with SPMHD, ideal MHD
was assumed. But the assumption of ideality is generally
not correct for the star formation processes because interstellar gas is
partially ionized and several magnetic diffusion processes (e.g., Ohmic dissipation,
Hall effect, and ambipolar diffusion) play roles. Especially, Ohmic
dissipation is effective at high density region ($\rho \gtrsim 10^{-12} \cm$)
and formation and evolution of  circumstellar disks, protostars and
jets are significantly affected by Ohmic dissipation
\citep[see, e.g.,][]{2006ApJ...647L.151M}.

To investigate the magnetic field in the intra-cluster medium of galaxy
clusters, SPMHD simulations with Ohmic dissipation were performed by \citet{2011MNRAS.418.2234B}.
But they only considered the spatially constant magnetic
resistivity. This assumption is generally not verified for protostellar
collapse simulations because the resistivity has large spatial variation
according to the gas density and temperature. Therefore, an Ohmic
dissipation scheme for SPMHD which includes the effect of the spatially-varying
resistivity is desired to investigate formation and evolution of protostars,
disks and jets.

In this paper, we  propose a new explicit scheme for Ohmic
dissipation with SPMHD. In section 2, we describe our SPH discretization for
Ohmic dissipation and time-stepping method. We present the results of
several numerical tests in section 3. Finally, we summarize our results
in section 4.

\section{Explicit Scheme}
\subsection{Discretization}
The induction equation with Ohmic dissipation is 
\begin{eqnarray}
\label{induction_eq}
\frac{d(\bm{B}/\rho)}{d t}=\frac{\bm{B}}{\rho}\cdot \nabla \bm{v} -\frac{1}{\rho}\nabla \times (\eta \nabla \times
\bm{B}),
\end{eqnarray}
where $\rho,~\bm{B},~\bm{v},~\eta$ denote the density, magnetic field,
velocity,  and
resistivity, respectively. Equation (\ref{induction_eq}) is solved by an
operator splitting approach and we focus on the solution of the second
term on the right-hand side.
The equation of Ohmic dissipation is given as
\begin{eqnarray}
\label{basic_eq}
\frac{d (\bm{B}/\rho)}{d t}=-\frac{1}{\rho}\nabla \times (\eta \nabla \times
\bm{B}).
\end{eqnarray}
Equation (\ref{basic_eq}) is written as,
\begin{eqnarray}
\label{eq0}
\frac{d (B_\mu/\rho)}{d t}=-\frac{1}{\rho} \{ \partial_\nu ( \eta
\partial_\mu B_\nu - \eta
\partial_\nu B_\mu ) \} \equiv \frac{1}{\rho} \partial_\nu F_{\mu\nu}.
\end{eqnarray}
Here, we used Greek letter, $\mu,~\nu$ to denote the components of
vector and we used Einstein summation convention. 
There are several choices for the discretization of the $(\nabla \cdot \bm{F})/\rho$.
In this study, we adopted the following discretization.
Discretization form of  $(\nabla \cdot \bm{F})/\rho$ of i-th particle is
\begin{eqnarray}
\label{final_term}
\left(\frac{1}{\rho}\partial_\nu F_{\mu\nu} \right)_i=
-\int \frac{\partial_\nu (\eta \partial_\mu
B_\nu - \eta \partial_\nu B_\mu)}{\rho}W_i d^3
 r  = \int \eta ( \partial_\mu B_\nu - \partial_\nu B_\mu)
 \partial_\nu \left(\frac{ W_i}{\rho} \right)d^3 r \nonumber \\ 
\sim -\sum_j m_j \left\{ \frac{\eta_i(\partial_\mu B_\nu -\partial_\nu B_\mu)_i}{\rho_i^2}+\frac{\eta_j(\partial_\mu
 B_\nu -\partial_\nu B_\mu)_j}{\rho_j^2} \right\} \partial_\nu W_{ij}.
\end{eqnarray}
Here, we used Latin letter, $i,~j$  to denote the particle number and
 $W_i=W[\bm{x}-\bm{x_i},h(\bm{x})]$ and
 $W_{ij}=W[\bm{x_i}-\bm{x_j},\bar{h}_{ij}]$, where we adopted the mean smoothing
length as $\bar{h}_{ij}=(h_i+h_j)/2$.

We also investigated the following formula for $(\nabla \cdot \bm{F})/\rho$, 
\begin{eqnarray}
\label{another_exp}
\left(\frac{1}{\rho}\partial_\nu F_{\mu\nu} \right)_i=
- \sum_j m_j\left[ 4
 \frac{1}{\rho_i\rho_j}\frac{(\eta_i\eta_j)}{(\eta_i+\eta_j)}\frac{(B_{\mu,j}-B_{\mu,i})}{|\bm{x_{ij}}|^2}(x_{\nu,i}-x_{\nu,j})\partial_\nu
 W_{ij}  +  \left\{\frac{\eta_i(\partial_\mu  B_\nu)_i}{\rho_i^2} +\frac{\eta_j(\partial_\mu B_\nu)_j}{\rho_j^2}\right\}\partial_\nu W_{ij}\right] .
\end{eqnarray}
%{\bf
The discretization of the first term on the right-hand side is suggested by \citet{1999JCoPh.148..227C}.
%}
The spatial resolution of this formula is slightly better than that of
equation (\ref{final_term})  but this introduces larger divergence
error because the discretizations of the derivative of the magnetic field 
and the volume factor are inconsistent between the first and the second term.   
Therefore, we adopted equation (\ref{final_term}).

There are also several choices for the gradient tensor of magnetic field. In
the following test calculations, we adopted
\begin{eqnarray}
\label{grad_tensor}
(\nabla \bm{B})_i=\frac{1}{\rho_i}\sum_j m_j(\bm{B}_j-\bm{B}_i)\nabla W[\bm{x_i}-\bm{x_j},h_i].
\end{eqnarray}
%{\bf
We use the cubic spline kernel of \citet{1985A&A...149..135M},
\begin{eqnarray}
W(r,h) = C_{\rm f} \left\{ \begin{array}{ll}
1 - \frac{3}{2} q^2 + \frac{3}{4} q^3, & 0 \le q < 1 \\
\frac{1}{4}(2-q)^3, & 1 \le q < 2 \\
0 & 2< q \end{array} \right.
\end{eqnarray}
where $q= r/h$ and $C_{\rm f}=\frac{1}{\pi h^3},~\frac{10}{7\pi h^2}$ for three and two dimensions, respectively.
The smoothing length of i-th particle is determined iteratively by the relation
\begin{eqnarray}
h_i=C_{\rm h}\left(\frac{m}{\rho_i}\right)^{1/d},
\end{eqnarray}
where $d$ is the dimension of the problem. $C_{\rm h}$ is a parameter and set to be $1.2$.

Although, we do not solve the energy equation in the following test calculations, it would be useful to derive
the SPH discretization of Ohmic dissipation term in the energy equation.
The energy equation of  Ohmic dissipation is given as
\begin{eqnarray}
\frac{D e}{Dt} \big\arrowvert_{{\rm Ohm}}=\frac{1}{\rho} \nabla \cdot \{\eta \bm{B} \times (\nabla \times \bm{B})\} = \frac{1}{\rho} \nabla \cdot  \left\{ \eta \nabla \left(\frac{\bm{B}^2}{2} \right) - \eta \bm{B}\cdot \nabla \bm{B} \right\} \equiv \frac{1}{\rho} \nabla \cdot \bm{S},
\end{eqnarray}
where $e=\frac{1}{2}{\bm v}^2+u+\frac{\bm{B}^2}{2\rho}$ is the specific total energy and $u=P/[(\gamma-1)\rho]$ is the
specific internal energy.
The discretization form of  $(\nabla \cdot \bm{S})/\rho$ of i-th particle is
\begin{eqnarray}
\label{energy_ohm}
\left(\frac{1}{\rho}\partial_\nu S_{\nu} \right)_i=
\int \frac{\partial_\nu S_\nu}{\rho}W_i d^3
 r  =- \int S_\nu
 \partial_\nu \left(\frac{ W_i}{\rho} \right)d^3 r \nonumber \\ 
\sim \sum_j m_j \left\{ \frac{(S_\nu)_i}{\rho_i^2}+\frac{(S_\nu)_j}{\rho_j^2} \right\} \partial_\nu W_{ij},
\end{eqnarray}
where $S_\nu$ is calculated as
\begin{eqnarray}
S_\nu=\eta \{\partial_\mu (\frac{B_\nu^2}{2}) - B_\mu (\partial_\mu B_\nu) \}=\eta \{B_\nu (\partial_\mu B_\nu) - B_\mu (\partial_\mu B_\nu) \},
\end{eqnarray}
and equation (\ref{grad_tensor}). The equation (\ref{energy_ohm}) is antisymmetric under particle exchange and
it is obvious that the error of the total energy is within machine epsilon by this discretization.

%}
\subsection{Time-stepping}
During the protostellar collapse, high-density region, $\rho \gtrsim
10^{-10} \cm$ appears.
In the high-density region, the timescale of Ohmic dissipation is shorter than the dynamical
timescale of the gas and the computational cost for Ohmic dissipation
becomes large. To reduce the computational cost,
we adopt the super-time-stepping method  (STS) proposed by
\citet{Alexiades96}. 
This method was used for Ohmic dissipation in \citet{2013ApJ...763....6T} and ambipolar diffusion in
\citet{2009ApJS..181..413C}.
In STS, Courant-Friedrich-Levy (CFL) stability condition is relaxed
by requiring the stability not at the end of each timestep but 
at the end of a cycle of $N_{{\rm sts}}$ steps.
Following \citet{Alexiades96}, we define a super timestep, $\Delta
T_{{\rm sts}} = \sum_{j=1}^{N_{{\rm sts}} } \tau_j$, where, $\tau_j$ is the sub-step and
given as,
\begin{eqnarray}
\tau_j=\Delta t_{{\rm exp}}[(1-\nu_{{\rm sts}})\cos(\frac{2j-1}{N_{{\rm sts}}}\frac{\pi}{2})+1+\nu_{{\rm sts}}]^{-1}.
\end{eqnarray}
Thus, the super timestep is
\begin{eqnarray}
\Delta T_{{\rm sts}}=\sum_{j=1}^{N{{\rm sts}}} \tau_j=\Delta
 t_{{\rm exp}}\frac{N_{{\rm sts}}}{2\nu_{{\rm sts}}^{1/2}}\frac{(1+\nu_{{\rm sts}}^{1/2})^{2N_{{\rm sts}}}
 -(1-\nu_{{\rm sts}}^{1/2})^{2N_{{\rm sts}}}}{(1+\nu_{{\rm sts}}^{1/2})^{2N_{{\rm sts}}}
 +(1-\nu_{{\rm sts}}^{1/2})^{2N_{{\rm sts}}}},
\end{eqnarray}
where $\Delta t_{{\rm exp}}$ is the explicit timestep for Ohmic dissipation
and we use $\Delta t_{{\rm exp}}=C_{{\rm CFL}}~h^2/2\eta$. Here, $C_{{\rm CFL}}$ is CFL
number and $h$ is the smoothing length. $\nu_{{\rm sts}}$ is a parameter that
controls the stability and the acceleration of the scheme. With smaller
$\nu_{{\rm sts}}$, the scheme becomes faster but unstable.
Optimal choice of $\nu_{{\rm sts}}$ depends on the problem and we investigate the optimal
choice for $\nu_{{\rm sts}}$ in section 3.
With STS, magnetic field is updated as
\begin{eqnarray}
(\frac{\bm{B}}{\rho})\big\arrowvert_{t+\Delta T_{{\rm sts}}}=(\frac{\bm{B}}{\rho})\big\arrowvert_{t}
+\sum_{j=1}^{N_{{\rm sts}}}\tau_j\frac{d(\bm{B}/\rho)}{dt}\big\arrowvert_{t+\sum_{k=0}^{j-1}\tau_k}.
\end{eqnarray}
For comparison, we also performed simulations with simple Euler method such as,
\begin{eqnarray}
(\frac{\bm{B}}{\rho})\big\arrowvert_{t+\Delta t_{{\rm exp}}}=(\frac{\bm{B}}{\rho})\big\arrowvert_{t}
+\Delta t_{{\rm exp}}\frac{d(\bm{B}/\rho)}{dt}\big\arrowvert_{t}.
\end{eqnarray}

\section{Numerical Tests}
\label{tests}
\subsection{Sinusoidal diffusion problem}
At first, we consider a simple problem in which sinusoidal magnetic
field diffuses with a constant resistivity.
The initial magnetic field is
\begin{eqnarray}
B_x=0,~B_y=0,~B_z(x)=\sin(2\pi x).
\end{eqnarray}
The resistivity is set to be $\eta=1$.
The computational domain is two dimensions and $x,y \in [-0.5,0.5]$. We imposed periodic
boundary conditions for each direction. 
We performed convergence tests by changing the timestep, $\Delta t_{{\rm exp}}$
and the smoothing length, $h$. As a measure of the error, we calculated 
$L_1$ norm of $B_z$ error, defined as
\begin{eqnarray}
L_1=\frac{1}{N_{{\rm tot}}}\sum_i^{N_{{\rm tot}}}|B_{{\rm ref},~z}(\bm{r}_i)-B_{z}(\bm{r}_i)|.
\end{eqnarray}
As  reference solutions, $B_{{\rm ref},z}$, we adopted the results with
$N_{{\rm tot}}=128^2 $, $\Delta t_{{\rm exp}}=6 \times 10^{-8}$ for the
convergence test of the timestep and $N_{{\rm tot}}=256^2$, $\Delta
t_{{\rm exp}}=1.5 \times 10^{-6}$ for the convergence test of the smoothing
length. For the calculations of both solutions, we used Euler method.
For STS, we adopted the value of $\nu_{{\rm sts}}=0.01,~N_{{\rm sts}}=5$.

Figure 1 shows the $L_1$ norms as a function of
timestep. We show both results of Euler method (solid) and STS (dashed). 
The horizontal axis is $\Delta t_{{\rm exp}}$ for Euler method  
and $\bar{\tau}$ for STS. Here, $\bar{\tau}$ is 
defined as $\bar{\tau}=\sum_j^{N_{{\rm sts}}}\tau_j/N_{{\rm sts}}$.
The figure shows that both schemes scale linearly and are first-order in
time.
This figure also shows that the error of STS is slightly
larger than Euler method at the same $\Delta t$. This means that, with the
same computational cost, the error of STS is slightly larger than Euler
method. This is simply because the error of STS is proportional not to
$\tau_j$ but to $\Delta T_{{\rm sts}}$.

Figure 2 shows the $L_1$ norms as a function of
smoothing length. The timestep is fixed to be $\Delta t_{{\rm exp}}=1.5\times
10^{-6}$ and Euler method is used.
The figure shows that the error is proportional to $h^2$ and it is
confirmed that our discretization is second-order in space.

\begin{figure}
\label{t_error_sinusoidal}
\includegraphics[width=60mm,angle=-90]{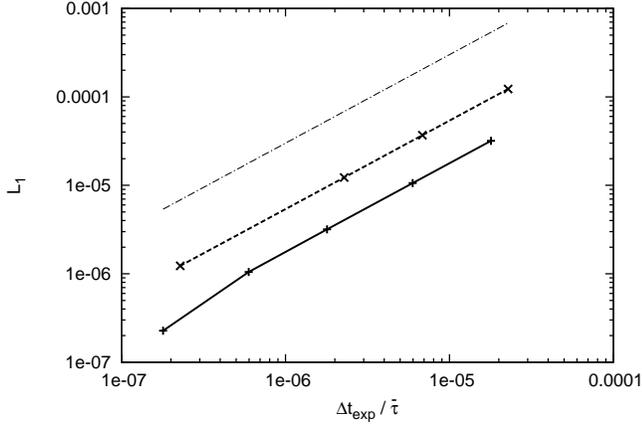}

\caption{$L_1$ norm of error as a function of timestep for the
 sinusoidal diffusion problem. Horizontal axis is $\Delta t_{{\rm exp}}$ for
 Euler method and $\bar{\tau}$ for STS. Solid line denotes the results
 with Euler method and dashed line denotes the results with STS. The
 dashed-dotted line is in proportion to $\Delta t$.
}

\end{figure}

\begin{figure}
\label{h_error_sinusoidal}
\includegraphics[width=60mm,angle=-90]{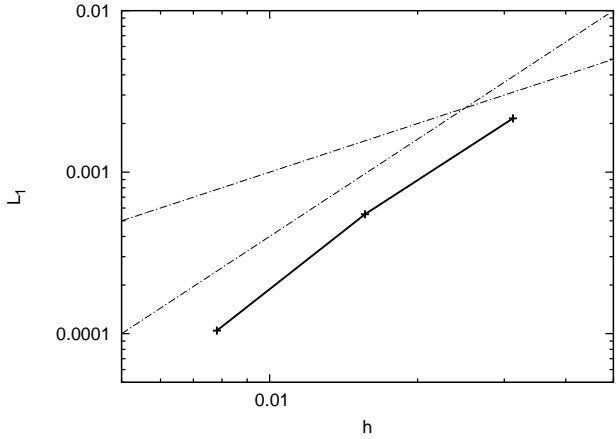}

\caption{
$L_1$ norm of error as a function of smoothing length for the
 sinusoidal diffusion problem. Solid line denotes the results
 with Euler method. The dashed-dotted lines are in proportion to $h$
 and $h^2$, respectively.
}
\end{figure}

\subsection{Gaussian diffusion problem}
Next, we consider magnetic diffusion of $B_z$ in Gaussian profile.
The initial profile of magnetic field is given as
\begin{eqnarray}
B_x=0,~B_y=0,~B_z(x,y)=\frac{1}{4\eta\pi t_0}\exp[-\frac{x^2+y^2}{4\eta
 t_0}],
\end{eqnarray}
where $t_0$ is the initial time and set to be unity. We set magnetic
resistivity as $\eta=1$.
The computational domain is two dimensions and  $x,y \in [-16,16]$. We impose periodic
boundary conditions for each direction.
The particle number for each direction is fixed to be 128.

Figure 3 shows the $L_1$ norm as a function of
timestep at $t=8$. Again, we considered both Euler method (solid) and STS (dashed). The reference
solution is the result with
$N_{{\rm tot}}= 128^2$, $\Delta t_{{\rm exp}}=1.4 \times 10^{-4}$ and Euler method.
The figure shows the same tendency of the results of sinusoidal
diffusion problem, i.e., Both schemes are first-order and the error of STS
method is slightly larger than Euler method at the same timestep.

To understand how the efficiency of STS depends on the parameters, 
we defined acceleration efficiency, $F=\Delta
T_{{\rm sts}}/(N_{{\rm sts}}\Delta t_{{\rm exp}})$ and plotted it in figure 4 with
$\nu_{{\rm sts}}=10^{-2},~10^{-3},~10^{-4}$. 
The figure shows that the maximum of acceleration efficiency is determined by given
$\nu_{{\rm sts}}$ and the maximum value is
$F_{max}=1/\sqrt{\nu_{{\rm sts}}}$. Therefore, small $\nu_{{\rm sts}}$ is preferable for the 
acceleration. But the small $\nu_{{\rm sts}}$ makes the scheme unstable.
This figure also shows the  efficiency saturates
around $N_{{\rm sts}}\sim 1/\sqrt{\nu_{{\rm sts}}}$. Therefore, optimal choice 
of $N_{{\rm sts}}$ is $\sim 1/\sqrt{\nu_{{\rm sts}}}$.

To seek the optimal choices of $\nu_{{\rm sts}}$ and $N_{{\rm sts}}$ for Ohmic
dissipation, we investigated the behavior of
the solutions at $t=8$ by changing the parameters. 
We choose the parameter sets as
$\nu_{{\rm sts}}=(10^{-2},10^{-3},10^{-4})$ and $N_{{\rm sts}}=5, 10$.
The CFL number is set to be $C_{{\rm CFL}}=0.3$ for all calculations.
The results at $y=0$ are shown in figure 5. In the figure, only 64
particles are plotted to make the results more visible.
The exact solution,
\begin{eqnarray}
B_z(x,y)=\frac{1}{4\eta\pi(t_0+t)}\exp[-\frac{x^2+y^2}{4\eta(t_0+t)}],
\end{eqnarray}
is also plotted.
The results with $N_{{\rm sts}}=5 $ (left panel) and
$(\nu_{{\rm sts}},N_{{\rm sts}})=(10^{-2},10)$ (circles in the right panel) are agree well with the
exact solution.
In the cases of $N_{{\rm sts}}=10$, as $\nu_{{\rm sts}}$ becomes small, the solution becomes distorted and the
result with $\nu_{{\rm sts}}=10^{-4}$ shows the significant
overshoot. This results shows that  $\nu_{{\rm sts}} \sim 0.01$ is preferable for the stability.
From figure 4, we can see that $F$ already saturate at $N_{{\rm sts}}\sim 5$ for $\nu_{{\rm sts}}=0.01$.
Therefore, we recommend $\nu_{{\rm sts}} \sim 0.01$ and $N_{{\rm sts}} \sim 5$ as
the optimal values of the parameters.

\begin{figure}
\label{t_error_gaussian}
\includegraphics[width=60mm,angle=-90]{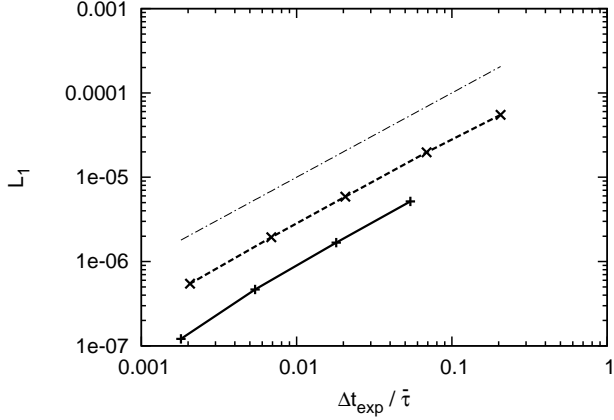}

\caption{
$L_1$ norm of error as a function of timestep for the
 Gaussian diffusion problem. Horizontal axis is $\Delta t_{{\rm exp}}$ for
 Euler method and $\bar{\tau}$ for STS. Solid line denotes the results
 with Euler method and dashed line denotes the results with STS. The
 dashed-dotted line is in proportion to $\Delta t$.
}
\end{figure}

\begin{figure}
\includegraphics[width=60mm,angle=-90]{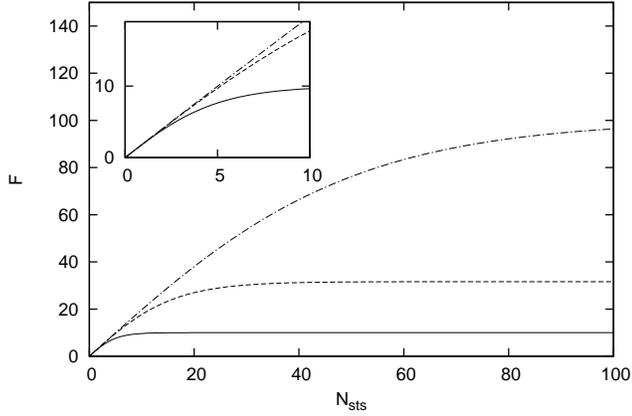}
\caption{ 
The acceleration efficiency, $F=\Delta
T_{{\rm sts}}/(N_{{\rm sts}}\Delta t_{{\rm exp}})$ as a function of $N_{{\rm sts}}$ for the case with
 $\nu_{{\rm sts}}=10^{-2}$ (solid), $10^{-3}$ (dashed), $10^{-4}$ (dashed-dotted). 
}
\end{figure}

\begin{figure}
\label{gaussian_nu_test}
\includegraphics[width=60mm,angle=-90]{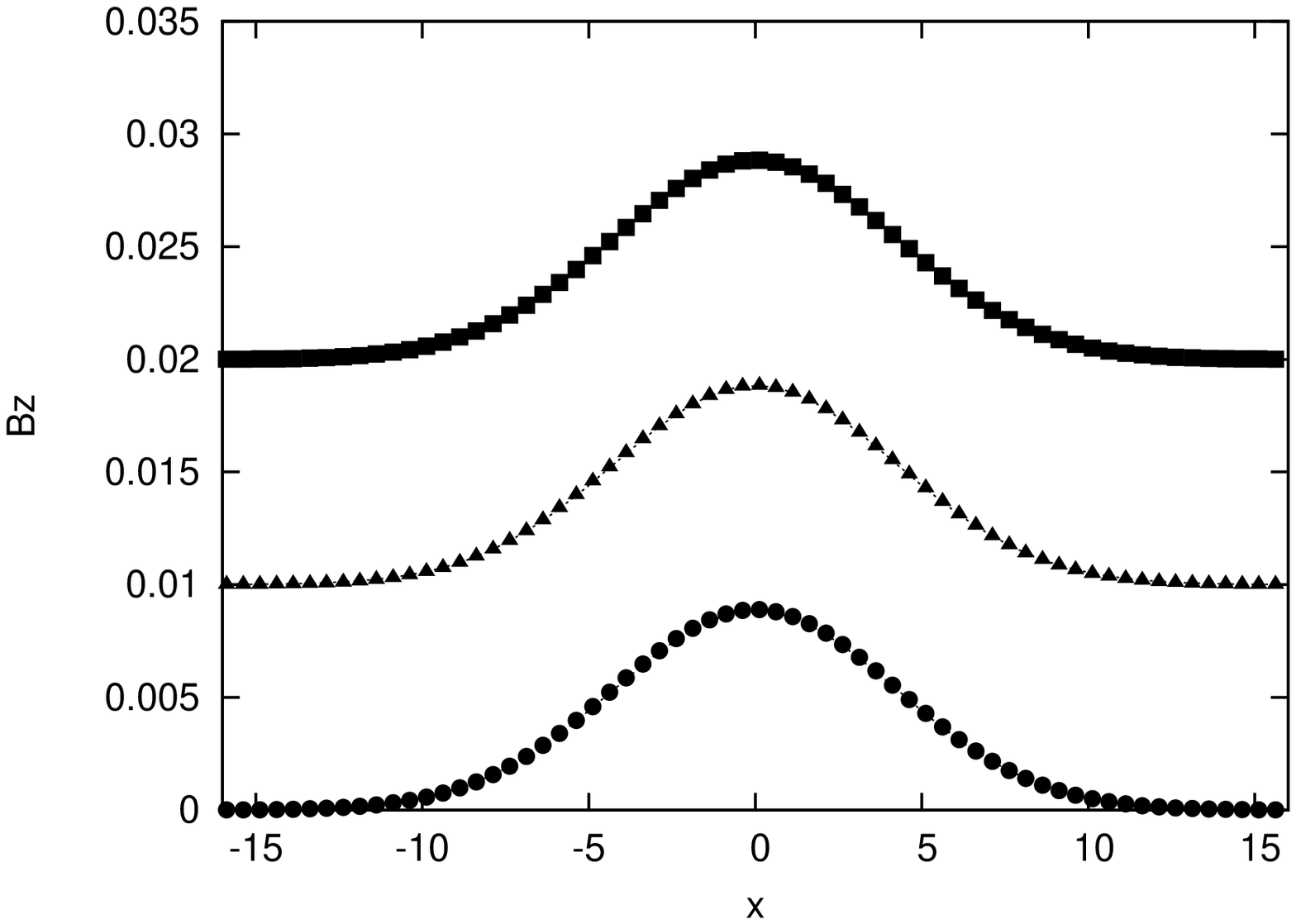}
\includegraphics[width=60mm,angle=-90]{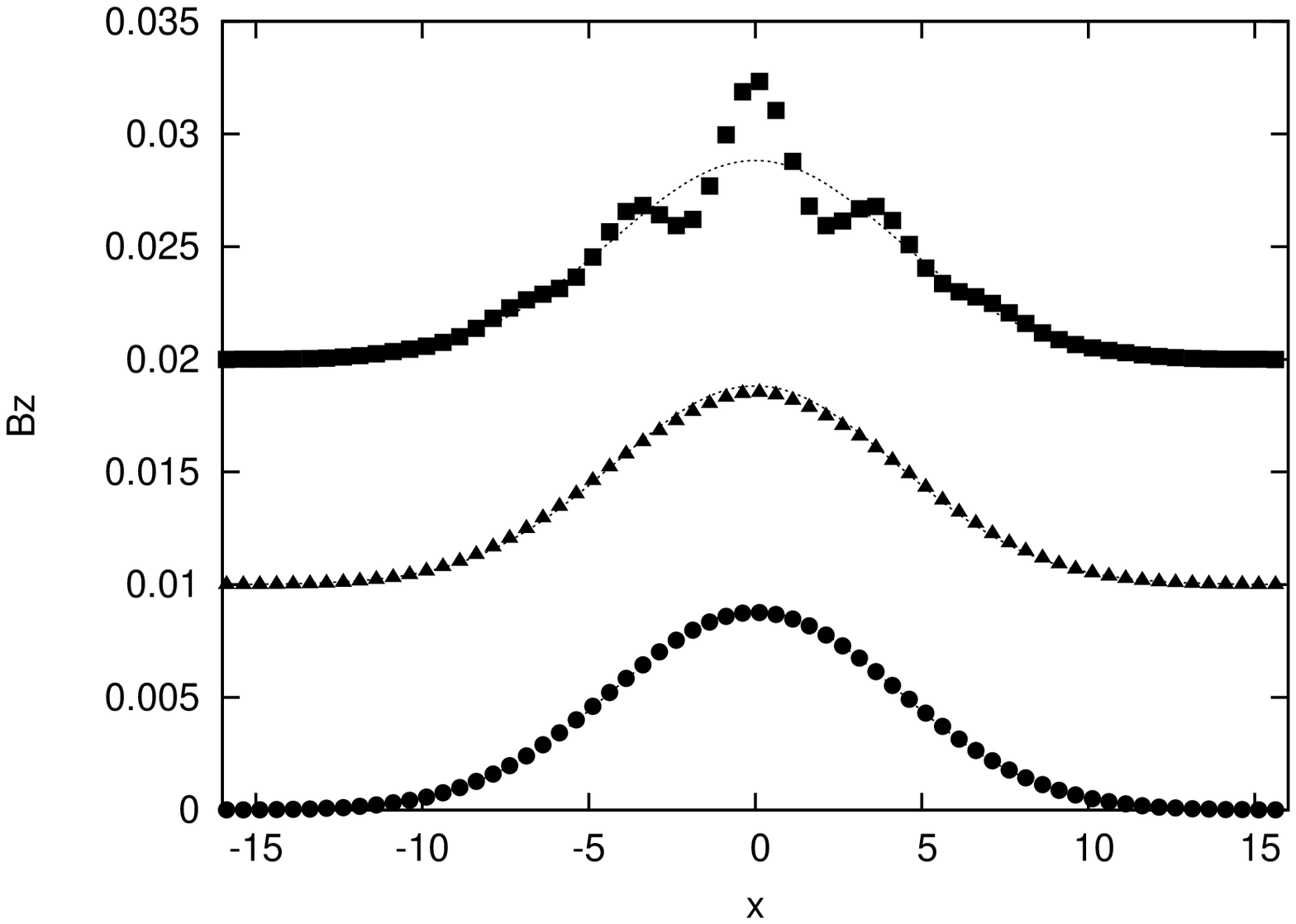}

\caption{ 
Profile of $B_z$ at $y=0$ plane at $t=8$ of Gaussian diffusion problem.
Circles, triangles, rectangles denote the results with  $\nu_{{\rm sts}}=$
 $10^{-2}$, $10^{-3}$, $10^{-4}$,
 respectively. Left and right panels show the results with $N_{{\rm sts}}=5$ and 
 $N_{{\rm sts}}=10$, respectively.
The dashed line denotes the exact solution at $t=8$. Each
 solution are offset from each other by 0.01 in the vertical direction
 to make the results more visible.
}

\end{figure}

\subsection{A test with spatially varying resistivity}
 \label{test3}
In this subsection, we consider the diffusion of $B_z$ in Gaussian profile with the
spatially varying resistivity.
The resistivity distribution is given as
\begin{eqnarray}
\eta(\bm{r})=\exp[-(x^2+y^2+z^2)],
\end{eqnarray}
and the initial magnetic field is 
\begin{eqnarray}
B_x=0,~B_y=0,~B_z(\bm{r})=\exp[-(x^2+y^2)].
\end{eqnarray}
The computational domain is three dimensions and $x,y,z\in [-3,3]$. We impose periodic
boundary conditions for each direction.

Figure 6 shows the contour maps of $B_x$ and  $B_z$ obtained at $t=1$ with Euler method
with $\Delta t_{{\rm exp}}=10^{-3}$ and STS with
$C_{{\rm CFL}}=0.3,~\nu_{{\rm sts}}=0.01,N_{{\rm sts}}=5$. The particle number of each
direction is 48.
The results are consistent with each other and also consistent with the 
results calculated with the grid-code \citep[see,][]{2011PASJ...63..317M}.
But the $B_z$ around the center is slightly overestimated with STS.

To confirm that our discretization is second-order in space with
the spatially varying resistivity, we show
the $L_1$ norm of $B_z$ at $t=1$ as a function of
smoothing length in figure 7. 
The solutions are obtained with Euler method and  $\Delta t_{{\rm exp}}=10^{-3}$.
The reference solution is the result with
$N_{{\rm tot}}= 96^3$ and $\Delta t_{{\rm exp}}=10^{-3}$.

The figure shows that the error is proportional to $h^2$ and it is
confirmed that our discretization is second-order in space.

\begin{figure}
\label{compare}
\includegraphics[width=90mm]{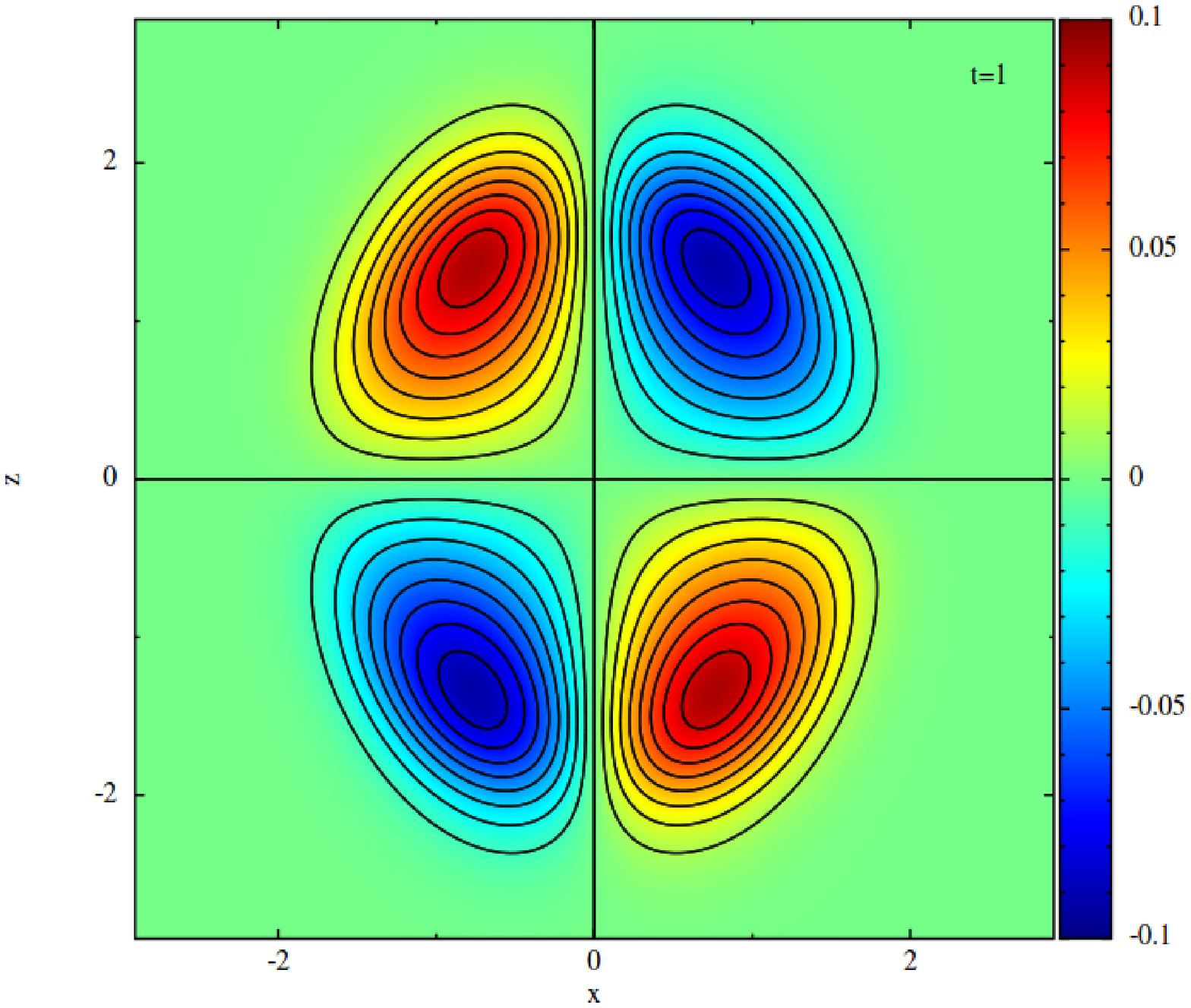}
\includegraphics[width=90mm]{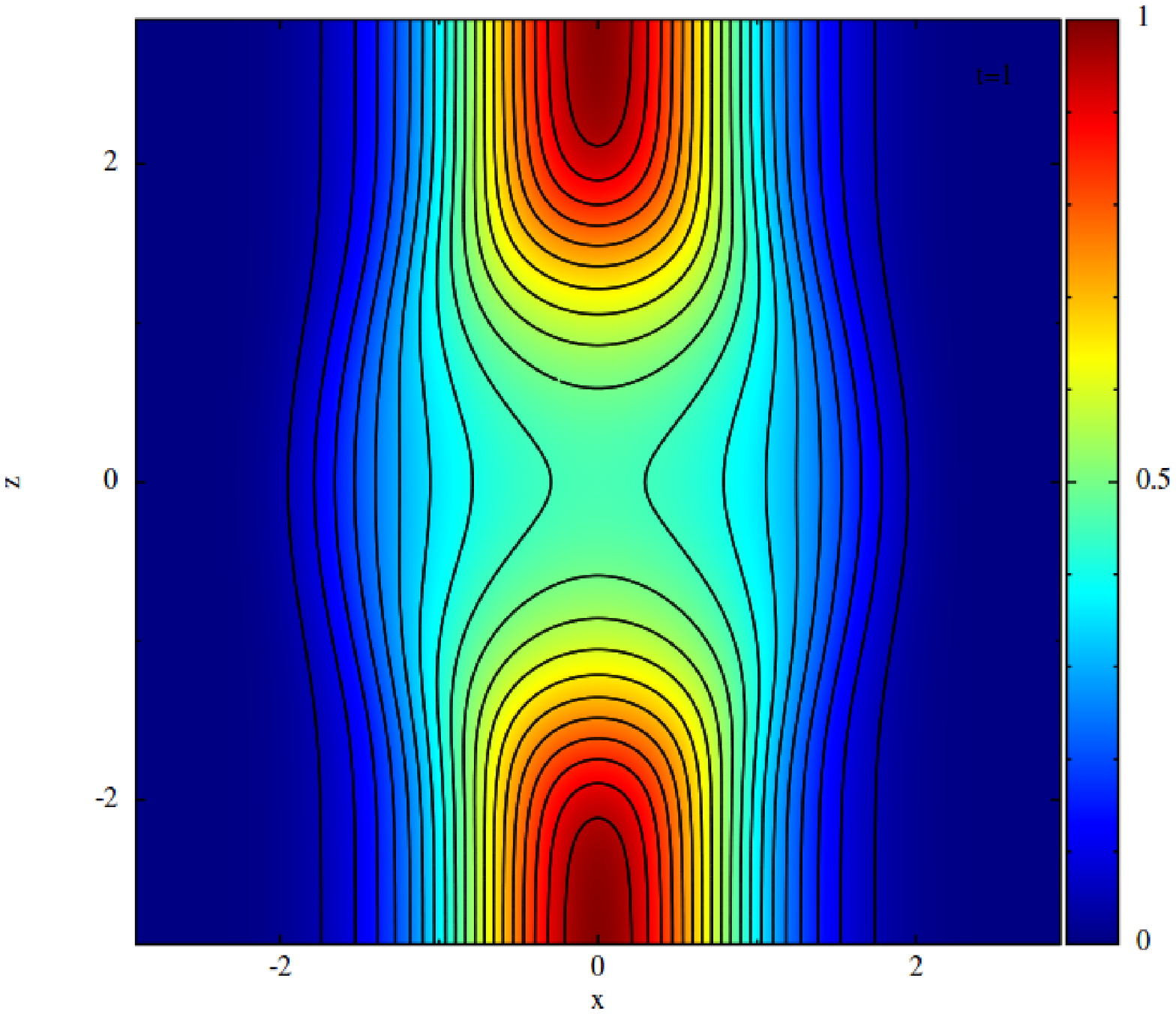}
\includegraphics[width=90mm]{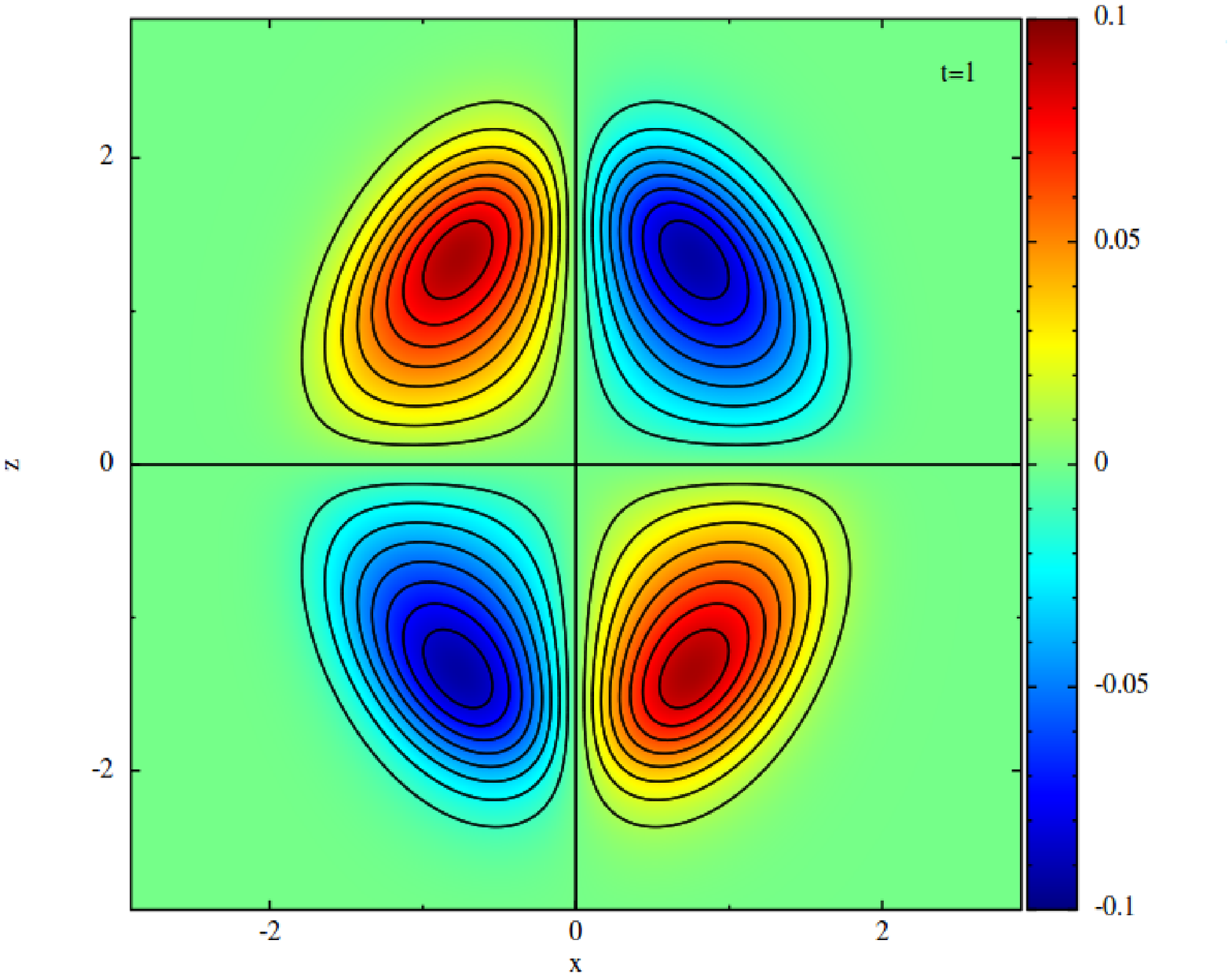}
\includegraphics[width=90mm]{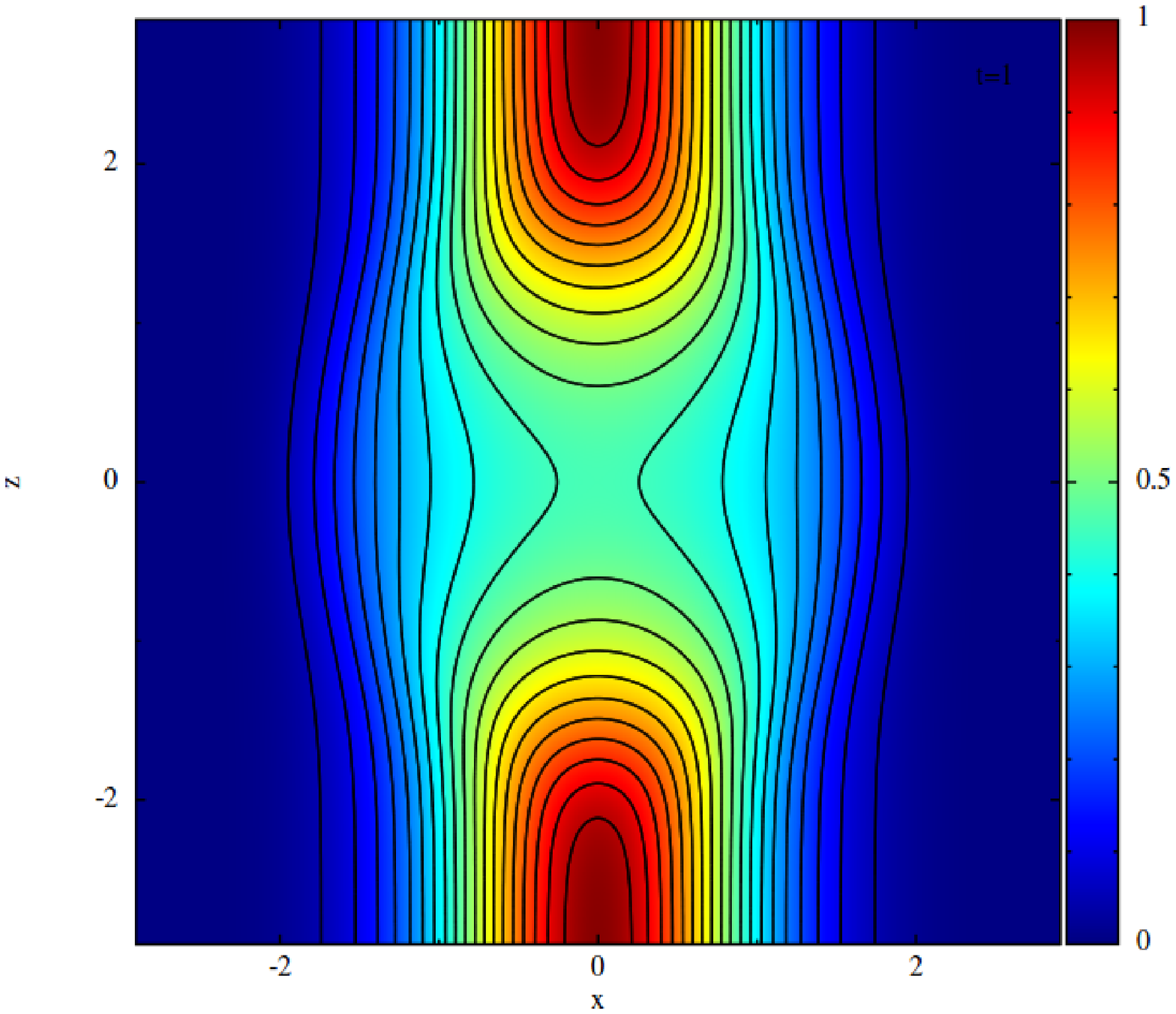}
\caption{
Magnetic field distributions of the 3D diffusion problem with
 spatially varying resistivity at $t=1$.  $y=0$ planes are shown.  
The  magnetic field is solved with Euler method and $\Delta t_{{\rm exp}}=10^{-3}$ for top
 panels and with STS and $C_{{\rm CFL}}=0.3,~\nu_{{\rm sts}}=0.01,N_{{\rm sts}}=5$ for
 bottom panels. Left panels show $B_x$ and contour levels are $B_x
 =-0.09,0.08, \cdots, 0.09$. Right panels show $B_z$ and contour levels are $B_z
 =0.05,0.1, \cdots, 0.95$.
}

\end{figure}

\begin{figure}
\label{h_error_3D}
\includegraphics[width=60mm,angle=-90]{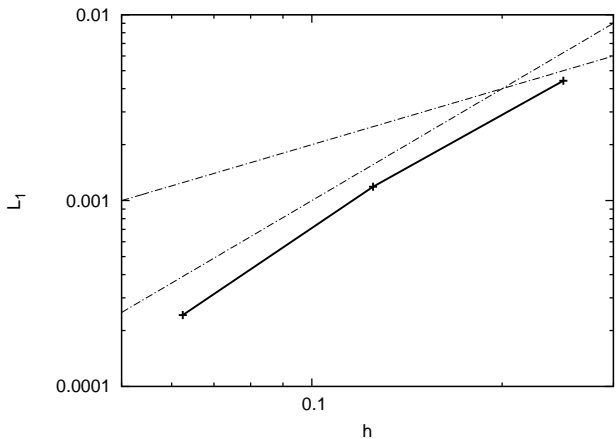}

\caption{
$L_1$ norm of error as a function of smoothing length for the
3D diffusion problem with spatially varying resistivity at $t=1$. Solid line denotes the results
 with Euler method and $\Delta t_{{\rm exp}}=10^{-3}$.
 The dashed-dotted lines are in proportion to $h$ and $h^2$ , respectively.
}

\end{figure}

\subsection{Gravitational collapse of magnetized cloud core}
 \label{gravity}
%{\bf
Finally, we consider the gravitational collapse of magnetized cloud core. The initial condition is as follow.
The initial molecular cloud core has a mass of $1~M_{\odot}$ and radius $R_c=2.7\times 10^4$ AU.
The free-fall time of the core is $2.4 \times 10^4$ years.  The core is rigidly rotating with
the angular velocity of $\Omega=1.8\times 10^{-13}~{\rm s^{-1}}$.
For the boundary condition, we fix the particles whose radius is larger than $2.6 \times 10^4$ AU.

We adopt a barotropic equation of state
\begin{eqnarray}
P = \left\{ \begin{array}{ll}
c_{\rm s}^{2} \rho (1+(\rho/\rho_{\rm c})^{2/5}), & \rho < \rho_{\rm d} \\
c_{\rm s}^{2}  \rho_{\rm c} (\rho_{\rm d}/\rho_{\rm c})^{7/5} (\rho/\rho_{\rm d})^{1.1} & \rho_{\rm d} \le \rho < \rho_{\rm e} \\
c_{\rm s}^{2}  \rho_{\rm c} (\rho_{\rm d}/\rho_{\rm c})^{7/5} (\rho_{\rm e}/\rho_{\rm d})^{1.1}(\rho/\rho_{\rm e})^{5/3} & \rho_{\rm e} \le \rho 
\end{array}\right.,
\end{eqnarray}
where $c_{\rm s}=190 {\rm m ~ s^{-1}}$, $\rho_{\rm c} = 4 \times 10^{-14}~\cm$, $\rho_{\rm d} = 4\times10^{-9}\cm$ and $\rho_{\rm e} = 4\times10^{-4}\cm$.
The initial magnetic field is parallel to z-axis with the magnitude of $B_{\rm z}=189~ \mu {\rm G}$ and the initial plasma beta 
is $\beta=2.5$. The cloud core is modeled with $5\times 10^6$ particles.

We use the GSPMHD scheme of \citet{2011MNRAS.418.1668I} with hyperbolic divergence cleaning method (Iwasaki \& Inutsuka in prep) 
to solve ideal MHD part and Barnes-Hut tree algorithm with opening angle $\theta=0.5$ for gravity part.
Ohmic dissipation is solved with present method.
We adopted the resistivity $\eta$ as 
\begin{eqnarray}
\eta = \dfrac{7.4 \times 10^5}{X_e}\sqrt{\dfrac{T}{10 {\rm K}}} \left[ 1-{\rm tanh}\left( \dfrac{n}{10^{12}\cmm}  \right)  \right] \, \, \, {\rm cm}^2\,{\rm s}^{-1},
\label{eq:etadef}
\end{eqnarray}
where $T$ and $n$ are the gas temperature and number density, and $X_e$ is the ionization degree of the gas and
\begin{eqnarray}
X_e =  5.7 \times 10^{-4} \left( \dfrac{n}{{\rm cm}^{-3}} \right)^{-1}.
\end{eqnarray} 
This model has the similar form to the model adopted in \citet{2007ApJ...670.1198M} but 
is artificially shifted to lower density to emphasize the effect of Ohmic dissipation in the first core. With our model,
Ohmic dissipation is effective at $10^{-13} \cm\lesssim \rho\lesssim 10^{-10} \cm$.

In figure 8, the magnetic energy of the central part (the region of $\rho>0.1 \rho_c$, where  $\rho_c$ is 
the central or maximum density of the cloud core. ) normalized by the thermal energy 
as a function of central density is shown.
The solid line and crosses show the results with STS and  Euler method, respectively. 
The result of ideal MHD is also shown with the dashed line for comparison. The parameters for STS are
$\nu_{{\rm sts}}=0.01,N_{{\rm sts}}=5$.

When the central density is small ($10^{-16}<\rho_c<10^{-14}~\cm$), Ohmic dissipation is ineffective and there is no
difference between resistive and ideal MHD models.  The magnetic energy of the resistive MHD models
begins to decrease at $\rho_c \sim10^{-13}~\cm$ and becomes more than three orders of magnitude smaller than the
ideal MHD model at  $\rho_c =10^{-10}~\cm$.
This figure also shows that the result with STS agree very well with that of the Euler method. Threrefore, 
STS is proved to be beneficial for the realistic star formation problems.

In figure 9, the density distributions at the center of the cloud when $\rho_c\sim5\times 10^{-3} \cm$
are shown,  The velocity field is shown with red arrows. 

In the ideal MHD model (left), the black thick line denotes the velocity contour of $|v_z|=0$.
This line clearly shows that the outflow forms at the center of the cloud.
On the other hand, in the resistive MHD model (right), the outflow does not form because of the
large resistivity in the first core. The structure of the first core is also very different from
the ideal MHD model because the magnetic braking is ineffective.
The detailed simulations and analysis of the formation and evolution of the 
outflow with ideal GSPMHD can be found in Iwasaki in prep.

\begin{figure}
\label{rho_Emag}
\includegraphics[width=60mm,angle=-90]{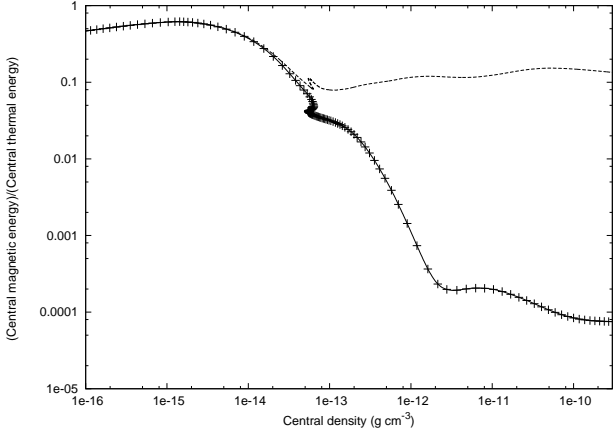}
\caption{
Central magnetic energy as a function of central density. 
The magnetic energy is normalized by the central thermal energy.
Solid line and crosses denote the results of the resistive MHD model
 with STS and Euler method, respectively.
 The dashed line denotes the result of the ideal MHD model.
}
\end{figure}

\begin{figure}
\label{outflow}
\includegraphics[width=70mm]{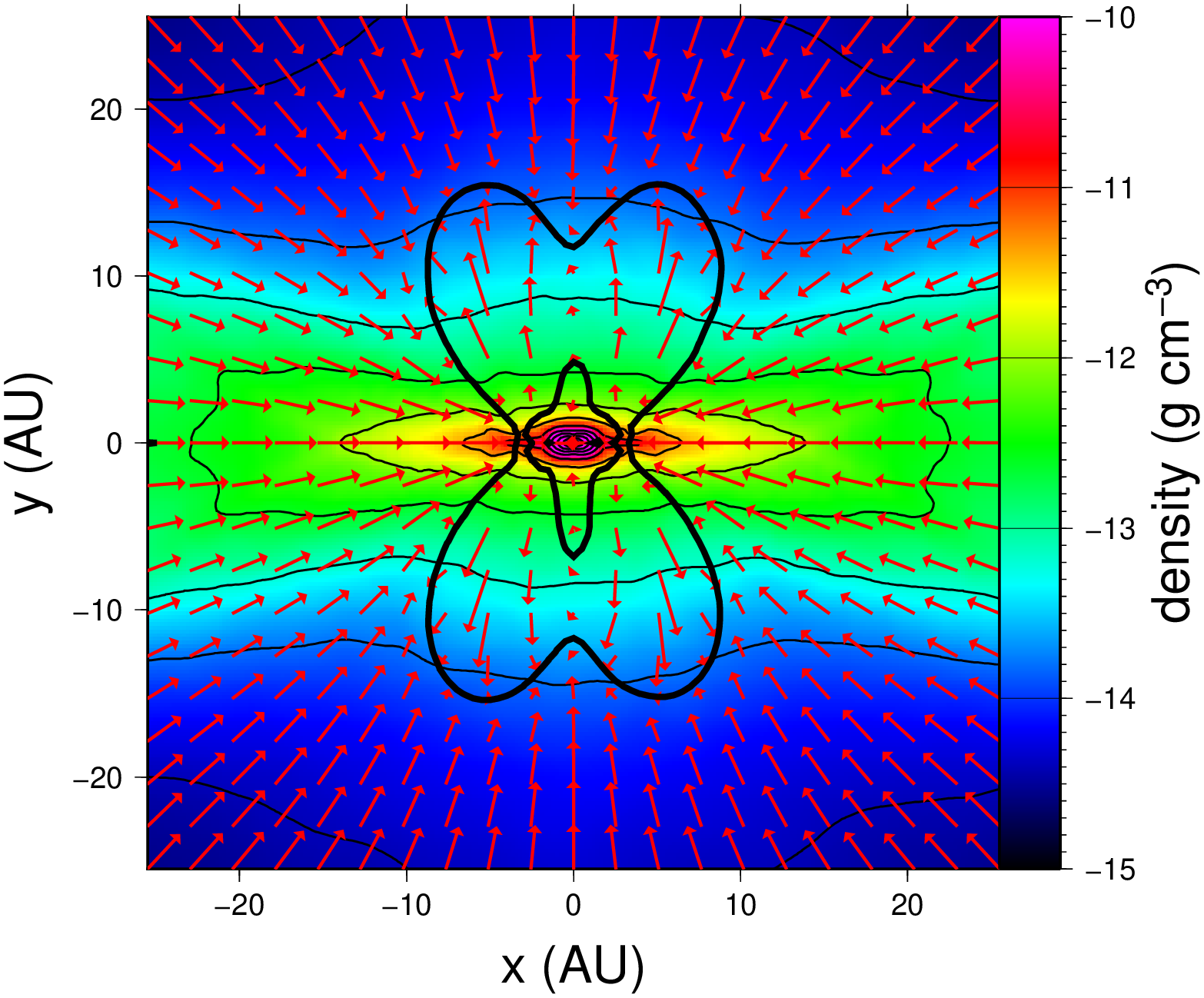}
\includegraphics[width=70mm]{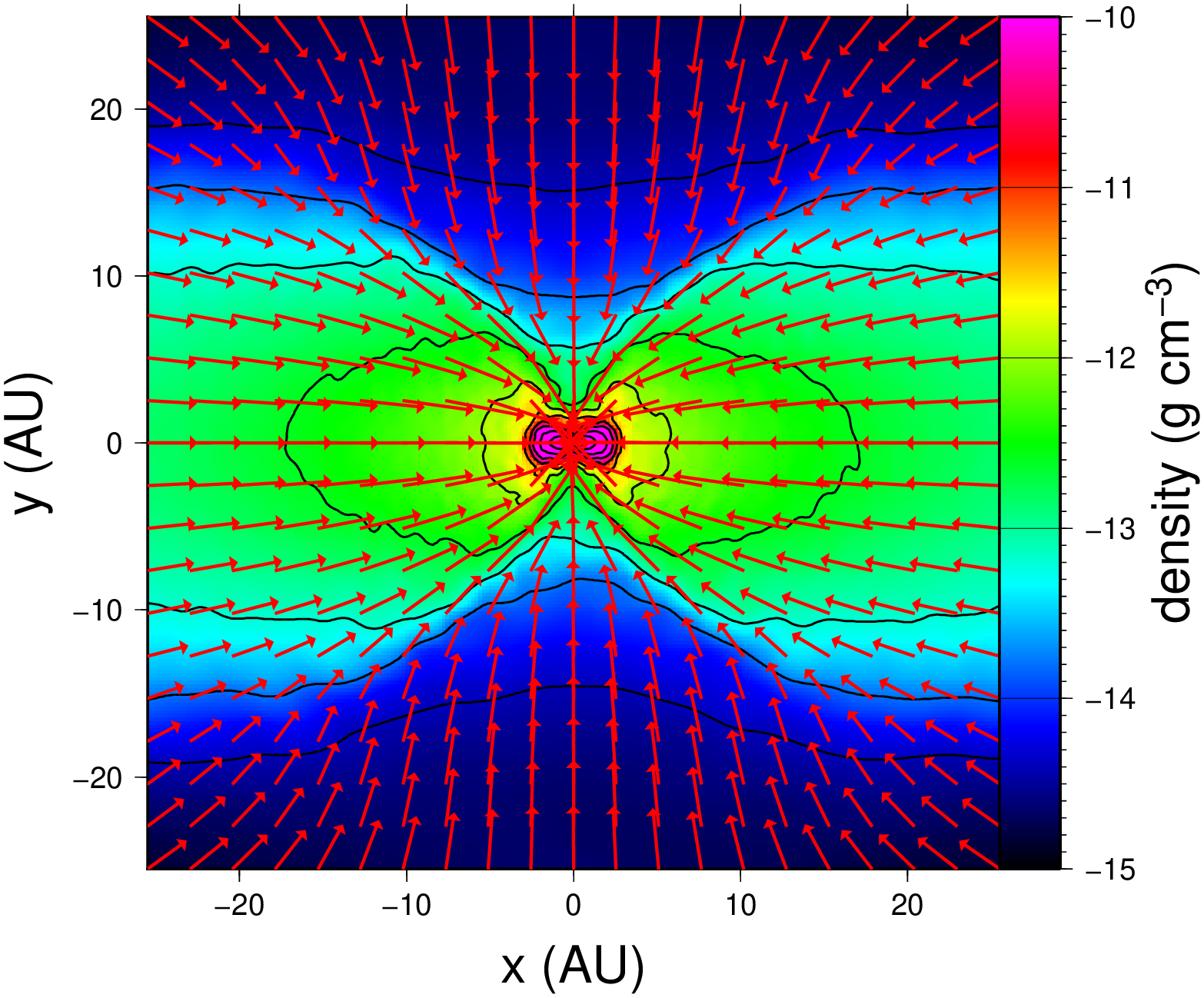}
\caption{
Density distributions at the center of the cloud of ideal MHD model (left) 
and resistive MHD model (right) when $\rho_c\sim5\times 10^{-3} \cm$.
$y=0$ planes are shown. The right panel is the result with STS.
The parameters for STS are
$\nu_{{\rm sts}}=0.01,N_{{\rm sts}}=5$.
The velocity field is shown with red arrows.
The thin black lines show the density contour and the thick black line in left panel shows the contour of $|v_z|=0$.
}
\end{figure}

%}

\section{Summary and Perspective}
\label{summary}
In this paper, we presented an explicit scheme for Ohmic dissipation
with smoothed particle magneto-hydrodynamics (SPMHD). We proposed a SPH
discretization of Ohmic dissipation term in the
induction equation. Ohmic dissipation part is solved with
super-time-stepping method (STS) which relaxes Courant-Friedrich-Levy (CFL) stability condition
requiring the numerical stability not at the end of each timestep but 
at the end of a cycle of $N_{{\rm sts}}$ steps.
Our scheme is second-order accurate in space
and first-order accurate in time. The scheme successfully solve
2D and 3D tests. 
Our scheme is simple and can be easily implemented to any SPMHD codes.

We showed that STS introduces slightly larger error compared to
Euler method if we fix the computational costs. This comes from the
fact that the error of STS is proportional not to $\tau$ but to $\Delta
T_{{\rm sts}}$.

We found that optimal choice of the parameters of STS for Ohmic dissipation of
SPMHD is $\nu_{{\rm sts}}\sim0.01$ and $N_{{\rm sts}}\sim 5$ and these values are
consistent with the values suggested by \citet{2013ApJ...763....6T}.

Our present scheme is only first-order accurate in time. 
Recently, \citet{2012MNRAS.422.2102M} suggest a method 
which extends STS to second-order accurate in
time. They applied this method to solve thermal conductivity.  It is
possible to solve Ohmic dissipation or other magnetic
diffusion with their method. 
Note that, however, the efficiency of acceleration of their method
is not so good as first-order STS at
small $N_{{\rm sts}}$ and large $N_{{\rm sts}}$ is required to achieve 
better acceleration.
We plan to improve accuracy in time of our scheme in future works.

\section *{Acknowledgments}
%We also thank anonymous referee for helpful comments.
We thank  T. Matsumoto, M. N. Machida, and T. Inoue for fruitful discussions. 
We also thank the referee, D. Price for helpful comments.
The snapshots were produced by SPLASH \citep{2007PASA...24..159P}.
The computations were performed on XC30 system at CfCA of
NAOJ and SR16000 at YITP in Kyoto University.
Y.T. and K.I. are financially supported by Research Fellowships of JSPS for Young Scientists.

\bibliography{article}

\begin{thebibliography}{23}
\expandafter\ifx\csname natexlab\endcsname\relax\def\natexlab#1{#1}\fi

\bibitem[{Alexiades, Amiez \& Gremaud(1996)Alexiades, Amiez, \&
  Gremaud}]{Alexiades96}
Alexiades V., Amiez G., Gremaud P.-A., 1996, Com. Num. Meth. Eng, 12, 12

\bibitem[{{Bate}(1998)}]{1998ApJ...508L..95B}
{Bate} M.~R., 1998, \apjl, 508, L95

\bibitem[{{Bonafede} {et~al}\mbox{.}(2011){Bonafede}, {Dolag}, {Stasyszyn},
  {Murante}, \& {Borgani}}]{2011MNRAS.418.2234B}
{Bonafede} A., {Dolag} K., {Stasyszyn} F., {Murante} G., {Borgani} S., 2011,
  \mnras, 418, 2234

\bibitem[{{Choi}, {Kim} \& {Wiita}(2009){Choi}, {Kim}, \&
  {Wiita}}]{2009ApJS..181..413C}
{Choi} E., {Kim} J., {Wiita} P.~J., 2009, \apjs, 181, 413

\bibitem[{{Dedner} {et~al}\mbox{.}(2002){Dedner}, {Kemm}, {Kr{\"o}ner}, {Munz},
  {Schnitzer}, \& {Wesenberg}}]{2002JCoPh.175..645D}
{Dedner} A., {Kemm} F., {Kr{\"o}ner} D., {Munz} C.-D., {Schnitzer} T.,
  {Wesenberg} M., 2002, Journal of Computational Physics, 175, 645

\bibitem[{{Inutsuka}(2002)}]{2002JCoPh.179..238I}
{Inutsuka} S.-I., 2002, Journal of Computational Physics, 179, 238

\bibitem[{{Inutsuka}, {Machida} \& {Matsumoto}(2010){Inutsuka}, {Machida}, \&
  {Matsumoto}}]{2010ApJ...718L..58I}
{Inutsuka} S.-i., {Machida} M.~N., {Matsumoto} T., 2010, \apjl, 718, L58

\bibitem[{{Iwasaki} \& {Inutsuka}(2011)}]{2011MNRAS.418.1668I}
{Iwasaki} K., {Inutsuka} S.-I., 2011, \mnras, 418, 1668

\bibitem[{{Machida}, {Inutsuka} \& {Matsumoto}(2006){Machida}, {Inutsuka}, \&
  {Matsumoto}}]{2006ApJ...647L.151M}
{Machida} M.~N., {Inutsuka} S.-i., {Matsumoto} T., 2006, \apjl, 647, L151

\bibitem[{{Machida}, {Inutsuka} \& {Matsumoto}(2011){Machida}, {Inutsuka}, \&
  {Matsumoto}}]{2011ApJ...729...42M}
---, 2011, \apj, 729, 42

\bibitem[{{Machida}, {Tomisaka} \& {Matsumoto}(2004){Machida}, {Tomisaka}, \&
  {Matsumoto}}]{2004MNRAS.348L...1M}
{Machida} M.~N., {Tomisaka} K., {Matsumoto} T., 2004, \mnras, 348, L1

\bibitem[{{Matsumoto}(2007)}]{2007PASJ...59..905M}
{Matsumoto} T., 2007, \pasj, 59, 905

\bibitem[{{Matsumoto}(2011)}]{2011PASJ...63..317M}
---, 2011, \pasj, 63, 317

\bibitem[{{Matsumoto} \& {Tomisaka}(2004)}]{2004ApJ...616..266M}
{Matsumoto} T., {Tomisaka} K., 2004, \apj, 616, 266

\bibitem[{{Meyer}, {Balsara} \& {Aslam}(2012){Meyer}, {Balsara}, \&
  {Aslam}}]{2012MNRAS.422.2102M}
{Meyer} C.~D., {Balsara} D.~S., {Aslam} T.~D., 2012, \mnras, 422, 2102

\bibitem[{{Price} \& {Monaghan}(2005)}]{2005MNRAS.364..384P}
{Price} D.~J., {Monaghan} J.~J., 2005, \mnras, 364, 384

\bibitem[{{Price} \& {Monaghan}(2007)}]{2007MNRAS.374.1347P}
---, 2007, \mnras, 374, 1347

\bibitem[{{Price}, {Tricco} \& {Bate}(2012){Price}, {Tricco}, \&
  {Bate}}]{2012MNRAS.423L..45P}
{Price} D.~J., {Tricco} T.~S., {Bate} M.~R., 2012, \mnras, 423, L45

\bibitem[{{Stamatellos}, {Whitworth} \& {Hubber}(2011){Stamatellos},
  {Whitworth}, \& {Hubber}}]{2011ApJ...730...32S}
{Stamatellos} D., {Whitworth} A.~P., {Hubber} D.~A., 2011, \apj, 730, 32

\bibitem[{{Tomida} {et~al}\mbox{.}(2013){Tomida}, {Tomisaka}, {Matsumoto},
  {Hori}, {Okuzumi}, {Machida}, \& {Saigo}}]{2013ApJ...763....6T}
{Tomida} K., {Tomisaka} K., {Matsumoto} T., {Hori} Y., {Okuzumi} S., {Machida}
  M.~N., {Saigo} K., 2013, \apj, 763, 6

\bibitem[{{Tricco} \& {Price}(2012)}]{2012JCoPh.231.7214T}
{Tricco} T.~S., {Price} D.~J., 2012, Journal of Computational Physics, 231,
  7214

\bibitem[{{Tsukamoto} \& {Machida}(2011)}]{2011MNRAS.416..591T}
{Tsukamoto} Y., {Machida} M.~N., 2011, \mnras, 416, 591

\bibitem[{{Tsukamoto} \& {Machida}(2013)}]{2013MNRAS.428.1321T}
---, 2013, \mnras, 428, 1321

\end{thebibliography}

\end{document}